\documentclass[12pt]{iopart}
\usepackage{graphics}
\usepackage{bm}
\usepackage{amssymb}
\usepackage{amsfonts}
\begin{document}
\title{Conditional purity and quantum correlation measures in two qubit mixed states}

\author{L.\ Reb\'on$^{1,2}$, R.\ Rossignoli$^{1,3}$,  J.J.M.\ Varga$^{4}$, N.\ Gigena$^{1}$,
N.\ Canosa$^{1}$,   C.\ Iemmi$^{4}$,  S.\ Ledesma$^{4}$}

\address{$^{1}$ Instituto de F\'isica de La Plata, Departamento de
F\'isica, CONICET, Universidad Nacional de La Plata, C.C. 67, 1900
La Plata, Argentina. $^{2}$ Institut f\"ur Laserphysik,
Universit\"at Hamburg, Luruper Chaussee 149, 22761 Hamburg, Germany.
$^{3}$ CICPBA, Argentina.$^4$ Departamento de F\'isica, FCEyN,
Universidad de Buenos Aires-CONICET, Buenos Aires 1428, Argentina.}
\ead{lrebon@PHYSnet.uni-hamburg.de}

\begin{abstract}
We analyze and show experimental results of the conditional purity,
the quantum discord and other related measures of quantum
correlation in mixed two-qubit states constructed from a pair of
photons in identical polarization states. The considered states are
relevant for  the description of spin pair states in interacting
spin chains in a transverse magnetic field. We derive clean
analytical expressions for the conditional local  purity and other
correlation measures obtained as a result of a remote local
projective measurement, which are fully verified by the experimental
results. A simple exact expression for the quantum discord of these
states in terms of the maximum conditional purity is also derived.
\end{abstract}
\maketitle

\section{Introduction}
Quantum correlations are at the heart of quantum information theory,
constituting  one of the key features that distinguishes quantum
from classical composite systems. They are recognized as the
essential resource that enables several quantum information
processing schemes \cite{NC.00}, such as quantum teleportation
\cite{Be.93} and  quantum algorithms exhibiting an exponential
speed-up over their classical counterparts \cite{Vi.03,RB.01}.
Although quantum entanglement \cite{Sch.95,RW.89} is the strongest
type of quantum correlations, it does not  encompass  all
non-classical correlations  that can be exhibited by mixed states of
composite systems. For such states, other measures of
quantum-like  correlations have been proposed \cite{KM.12}, starting with
quantum discord \cite{OZ.01,HV.01,Ve.03,Zu.03,KW.04,Ca.08},  which was then followed by 
other related measures \cite{KM.12,DVB.10,SKB.11,RCC.10}. The
presence of discord-like correlations has been shown to be 
important in various quantum information processes \cite{KM.12}, 
including quantum state merging \cite{Ca.11}, quantum state discrimination \cite{Roa2011}, quantum
cryptography \cite{Pi.14}, quantum metrology \cite{G.13} and quantum protocols 
 \cite{BP.14}. Accordingly, several studies and verifications of these correlations 
in distinct contexts have been made \cite{KM.12,CRC.10,WW.11,AB.12,HH.14,QZZ.15}. 
Furthermore, a quantum resource theory based on quantum coherence has recently been developed \cite{Ba.14,St.15} 
and some quantifiers of this resource have been directly related to measures of quantum discord \cite{Yxgs.15,Ma.16}.

Quantum discord for a bipartite system can be defined \cite{OZ.01}
as the minimum difference between two distinct quantum generalizations
of the classical conditional von Neumann entropy, one being the
standard formal extension of the classical expression while the
other one, introduced in \cite{OZ.01}, involves a local measurement
on one of the constituents. The latter generalization measures the
average conditional mixedness of the unmeasured component ($A$)
after the local measurement on the other component ($B$) and  is
always a positive quantity, which is smaller or at least never
larger than the original marginal entropy $S(A)$. Moreover, its minimum over all local measurements at $B$ 
is exactly the entanglement of formation between $A$ and a third system $C$ purifying the whole system \cite{KW.04}. 
This fundamental relation has enabled, in particular,  the connection with quantum state merging, a quantum information task where 
in a pure tripartite system $ABC$, $A$ transfers her state to $C$ through classical communication and 
shared entanglement. The minimal entanglement consumed in such process is given by the quantum discord between $A$ and $B$ 
(with measurements at $B$). The  measurement dependent conditional entropy  has been recently extended to more general entropic forms \cite{GR.14}, 
where a similar relation with the corresponding generalized entanglement of formation holds. 
In particular, this generalization allows the use of simple
forms like the linear entropy, which can be more easily evaluated and enables a clean
analytical solution of the associated optimization problem (i.e.,
that of selecting the local measurement which minimizes the
conditional entropy) in general qudit-qubit states \cite{GR.14}. Moreover, such entropy 
is directly related to the purity, an experimentally
accessible quantity whose determination does not require a full state tomography
\cite{F.02,B.04}, and which in the case of a qubit is formally related to the classical degree of polarization 
\cite{GJ.12}. 

The aim of this work is to analyze and experimentally obtain the
conditional purity, the quantum discord and other related measures
of discord-like correlations \cite{DVB.10, RCC.10} in a particular
class of mixed states which can be faithfully simulated through
photonic quantum systems and linear optics
\cite{Mataloni2014,Dakic2014}. These states, which are mixtures of
non-orthogonal aligned states, arise naturally in different
contexts, in particular as reduced pair states in the exact ground
state of spin $1/2$ chains with anisotropic $XY$ or $XYZ$ couplings,
in the immediate vicinity of the factorizing magnetic field
\cite{Ku.82,RCM.08,CRM.10}. In these chains they can also provide a
basic description of reduced pair states in mean-field symmetry-breaking
phases ($B<B_c$) \cite{CRC.10}.

For such states, described in section \ref{IIA}, we first
derive in  \ref{IIB}--\ref{IIC} simple  analytical expressions for
the conditional reduced state and its purity after a local
measurement on one of the qubits, including the average conditional
purity and its maximum among all possible local projective
measurements. We then derive in sec.\ \ref{IID} a simple closed
analytical expression for the quantum discord of these states in
terms of the maximum conditional purity. Note that the computation of the
quantum discord for general states  is difficult due the associated minimization,
having recently been shown to be an NP-complete problem \cite{YH.14}.
Additionally, in \ref{IIE} we present expressions for
the global post-measurement purity  and its minimizing measurement,
which allows to evaluate the associated information deficit and the
geometric discord. These quantities are also analyzed and compared.
 We have then experimentally tested these theoretical results using
polarization-encoded photonic qubits
that arise from a source emitting a pair of photons in the same
polarization state, which enables to reproduce the mixed two qubit
states. A description of the experimental setup used to prepare the
desired state and perform the local measurements is given at the
beginning of section \ref{IIIA}, with \ref{IIIB} devoted to present
the experimental results and their comparison with the theoretical
predictions. Conclusions are finally given in \ref{IV}.

\section{Theory}
\subsection{Initial state\label{IIA}}
We consider the symmetric two qubit mixed state
\begin{equation}
\rho_{AB}=p|\theta\theta\rangle\langle\theta\theta|+q|\!-\!\theta\!-\!\theta\rangle
\langle\!-\theta\!-\!\theta|\,,\label{state}
\end{equation}
where $|\theta\theta\rangle\langle\theta\theta|
\equiv|\theta\rangle\langle\theta|\otimes|\theta\rangle\langle\theta|$, with
\begin{equation}|\pm\theta\rangle=
{\textstyle\cos\frac{\theta}{2}|0\rangle\pm\sin\frac{\theta}{2}|1\rangle}\,,
\end{equation}
pure single qubit states forming angles $\pm\theta$ with the $z$
axis on the Bloch sphere,
and $p\in[0,1]$,  $q=1-p$ are the probabilities of
preparing both qubits in the states $|\theta\rangle$ and
$|\!-\!\theta\rangle$ respectively.
Any rank $2$ mixture of the form
$\rho_{AB}=p|\Omega\Omega\rangle\langle\Omega\Omega|+
q|\Omega'\Omega'\rangle\langle\Omega'\Omega'|$,
with
$|\Omega\rangle=\cos\frac{\theta}{2}|0\rangle+e^{i\phi}\sin\frac{\theta}{2}|1\rangle$
a general qubit state, can be rewritten at once in the form
(\ref{state}) by choosing a new $z$ axis in the Bloch sphere halfway
between the directions $\Omega=(\theta,\phi)$ and
$\Omega'=(\theta',\phi')$ (and the $x$ axis in the plane determined
by them). Moreover, {\it any} mixture
$\rho_{AB}=p|\Omega_1\Omega_2\rangle\langle \Omega_1\Omega_2|+
q|\Omega'_1\Omega'_2\rangle\langle\Omega'_1\Omega'_2|$,
where the angle between $\Omega'_2$ and $\Omega_2$ is identical with
that between $\Omega'_1$ and $\Omega_1$,
can be also brought to the form (\ref{state}) by applying
local rotations on one of the qubits
that  shift $\Omega_2$ to $\Omega_1$ and  $\Omega'_2$ to $\Omega'_1$.
These rotations will not affect correlation measures.

Mixed states of the form (\ref{state}) can arise in different
contexts. For instance, they emerge naturally as reduced two-spin states in the
ground state  of ferromagnetic-type spin $1/2$ arrays with
anisotropic $XY$ or $XYZ$ couplings
in the immediate vicinity  of the transverse factorizing magnetic field
\cite{CRC.10,RCM.08,CRM.10}, where the exact ground state becomes
two-fold degenerate, being an arbitrary linear combination of uniform
completely separable states, i.e.,
\begin{equation}|GS\rangle=\alpha|\theta\theta\ldots\rangle+
\beta |\!-\!\theta\!-\!\theta\ldots\rangle\label{GS}\,,\end{equation}
where $\theta$ is determined by the coupling
anisotropy \cite{RCM.08} (assumed constant for all coupled pairs).
The state (\ref{GS}) leads to the reduced  two-spin state (\ref{state})
with $p=|\alpha^2|$, $q=|\beta^2|$ for {\it any} pair $i\neq j$, after
tracing out the remaining qubits
and neglecting the complementary overlap $\langle-\theta|\theta\rangle^{n-2}
=\cos^{n-2}\theta$, which decreases exponentially with increasing $n$ if $|\cos\theta|<1$.
And in the mean field approximation,  a reduced
state of the form (\ref{state}) with $p=q=1/2$ naturally arises in the whole parity breaking phase
after parity symmetry restoration \cite{CRC.10,RCM.08,CRM.10}, becoming exact at the factorizing point.

The states (\ref{state}) can also  be generated using a source emitting a
pair of photons in the same polarization state,  e.g.\ by spontaneous
parametric down-conversion (SPDC) produced in nonlinear crystal cut
for type I phase matching \cite{Burnham70,MandelBook} and linear
optics, such that
\begin{equation}|\pm\theta\rangle={\textstyle\cos\frac{\theta}{2}
|V\rangle\pm\sin\frac{\theta}{2}|H\rangle}
\label{statep}\end{equation} are linearly polarized states at angles
$\pm\theta/2$ with the vertical direction. Here
$|V\rangle\equiv|0\rangle,\,|H\rangle\equiv|1\rangle$ denote the
orthogonal linearly polarized states in the vertical and horizontal
directions respectively.

The purity of the state (\ref{state}) is given by
\begin{equation}P_{AB}={\rm Tr}\,\rho_{AB}^2=1-2pq(1-\cos^4\theta)\,,  \label{PAB}\end{equation}
and is an increasing function of the overlap $|\langle\!-\!\theta|\theta\rangle|=|\cos\theta|$.
Since $\rho_{AB}$ is a rank $2$ state, the purity (\ref{PAB}) completely determines its two non-zero eigenvalues,
\begin{eqnarray}\lambda^{\pm}_{AB}&=&{\textstyle\frac{1}{2}}[1\pm \sqrt{2P_{AB}-1}]\,,\label{lab}\end{eqnarray}
and hence, the value of any entropy
\begin{equation}S_f(\rho_{AB})={\rm Tr}\,f(\rho_{AB})=f(\lambda^+_{AB})+f(\lambda^-_{AB})\,,
 \label{Sf}\end{equation}
where $f$ is a concave function satisfying $f(0)=f(1)=0$ \cite{Wh.78,CR.02}.
In particular, the von Neumann entropy $S(\rho)$ corresponds
to $f(\rho)=-\rho\log_2\rho$, while the  linear entropy $S_2(\rho)$
 to $f(\rho)=-2\rho(1-\rho)$, in which case
\begin{equation}S_2(\rho_{AB})=2(1-{\rm Tr}\,\rho_{AB}^2)=2(1-P_{AB})\,.\label{6}\end{equation}
All entropies (\ref{Sf}) will be decreasing functions of $P_{AB}$,  vanishing iff $P_{AB}=1$.

The state (\ref{state}) is {\it separable},  i.e., a convex mixture of product
states \cite{RW.89}. Nonetheless, if $\theta\in(0,\pi/2)$
and $pq\neq 0$, it is not classically correlated, i.e., it is not diagonal in a standard or 
conditional product basis, having {\it entangled} eigenstates. 
It will then lead to a finite quantum discord (sec.\ \ref{IID}).

The reduced state of each of the qubits (or photons) is
\begin{eqnarray}\rho_B&=&{\rm Tr}_{A}\,\rho_{AB}=
p|\theta\rangle\langle\theta|+q|-\theta\rangle\langle-\theta|\label{rhoa}\\
&=&\frac{1}{2}\left(\begin{array}{c c}
 1+\cos\theta & (p-q)\sin\theta\\
 (p-q)\sin\theta & 1-\cos\theta
\label{mr}
\end{array}\right)
\end{eqnarray}
where (\ref{mr}) is the representation in
the standard basis $\{|V\rangle,|H\rangle\}$, and corresponds to a Bloch vector $\bm{r}_B=
{\rm Tr}\,\rho_B\,\bm{\sigma}=((p-q)\sin\theta,0,\cos\theta)$.
The {\it local} purity $P_A=P_B$ is then
\begin{equation}P_{B}={\rm Tr}\,\rho_B^2=1-2pq\sin^2\theta\label{pa}\,,\end{equation}
with the eigenvalues of $\rho_B$ given by $\lambda_B^{\pm}=(1\pm\sqrt{2P_B-1})/2$.
It is verified that $P_B\geq P_{AB}$, $\lambda^+_{B}\geq \lambda^+_{AB}$, in
agreement with the general majorization properties \cite{Bha.97}
$\rho_{AB}\prec\rho_{B(A)}$ valid for separable mixed states $\rho_{AB}$
\cite{NK.02,RC.03}.

 \subsection{Conditional local state and purity after remote local measurement \label{IIB}}
\begin{figure}[ht]
\centerline{\hspace*{-0.1cm}\scalebox{1.2}{\includegraphics{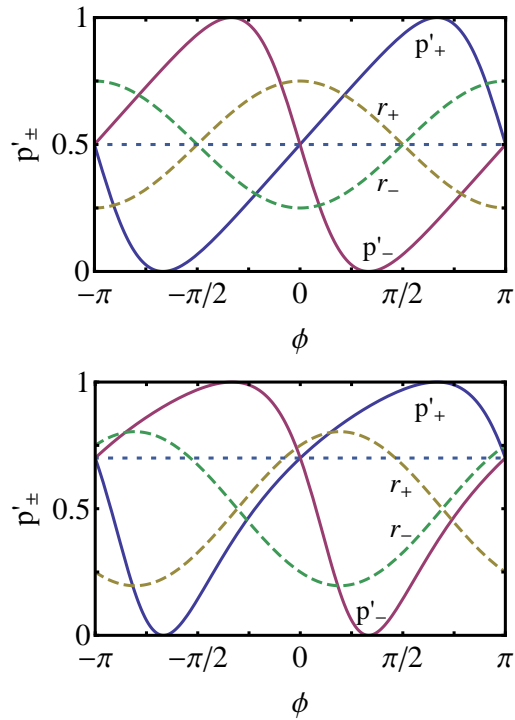}}}
\vspace{0cm} \caption{(Color online) The probabilities $p'_{\pm}$, Eq.\ (\ref{pac}),
of the conditional reduced states (\ref{rhoac}) of $A$ after a measurement with  result
$\pm$ at $B$ along angle $\phi$, for $\theta=\pi/3$ and initial values $p=0.5$ (top)
and $p=0.7$ (bottom), indicated by the horizontal dotted lines.
It is seen that $p'_{\pm}$ cover all values between $0$ and $1$, with $p'_+=1$ at $\phi=\pi-\theta$
and $0$ at $\phi=\theta-\pi$, while $p'_-=0$ at $\phi=\theta$ and $1$ at $\phi=-\theta$.
The dashed lines depict the probabilities $r_{\pm}$, Eq.\ (\ref{r}), of obtaining result
$\pm$ at $B$.} \label{f1}
\end{figure}

Let us consider now a projective polarization measurement on photon $B$,
defined by the orthogonal projectors
\begin{equation}
\Pi_+=|\phi\rangle\langle\phi|\,,\;\;\Pi_-=|\phi+\pi\rangle\langle\phi+\pi|
\,,\label{proj}\end{equation}
where
$|\phi\rangle={\textstyle\cos\frac{\phi}{2}|V\rangle+\sin\frac{\phi}{2}|H\rangle}$,
$|\phi+\pi\rangle=-{\textstyle\sin\frac{\phi}{2}|V\rangle+\cos\frac{\phi}{2}|H\rangle}$
and $\Pi_++\Pi_-=1$. This means projecting onto linearly polarized states at
angles $\phi/2$ and $\phi/2+\pi/2$ respectively.
 The probability of obtaining result $+$ or $-$ is
\begin{eqnarray}
r_{\pm}={\rm Tr}\,(\rho_{AB}\,I_A\otimes \Pi_{\pm})=
{\textstyle\frac{1}{2}}[1\pm p\cos(\phi-\theta)\pm q\cos(\phi+\theta)]\,.\label{r}\end{eqnarray}
(Obviously, a result ``$+$'' for measurement angle $\phi$ is equivalent to a
result ``$-$'' for measurement angle $\phi+\pi$).
As a function of $\phi$, $r_\pm$ is extremum for
\begin{equation}\tan\phi=(p-q)\tan\theta\,,\end{equation}
with $r_+$ maximum for $\phi=0$ if $p=q$ and $\phi$ between $0$ and $\theta$ if $p>q$
(Fig.\ \ref{f1}).

After a measurement at $B$ with known result,  the post-measurement
state of the unmeasured photon will have again the form
(\ref{rhoa}),  but with modified probabilities $p'_{\pm},q'_{\pm}$,
which depend on both the measurement angle $\phi$ and measurement
result $\pm$:
\begin{eqnarray}\rho_{A/B_{\pm}}&=&r_{\pm}^{-1}{\rm Tr}_B\,(\rho_{AB}\,I_{A}\otimes \Pi_{\pm})=
p'_{\pm}|\theta\rangle\langle\theta|+q'_{\pm}|-\theta\rangle\langle-\theta|\,,
\label{rhoac}\end{eqnarray}
with $q'_{\pm}=1-p'_{\pm}$ and
\begin{eqnarray}p'_{\pm}&=&p\frac{1\pm\cos(\theta-\phi)}{2r_\pm}\,.
\label{pac}\end{eqnarray}
It is of course verified that $r_+\rho_{A/B_+}+r_-\rho_{A/B_-}=\rho_A$,
i.e., that the post-measurement state at $A$ is unchanged if the result is unknown. It is also
seen that if $\phi=0$ (or $\phi=\pi$), $p'_{\pm}=p$, i.e., $\rho_{A/B_{\pm}}=\rho_A$
{\it irrespective} of the values of $\theta$, $p$ and the result of the measurement.
Such measurement then leaves  the marginal state of the unmeasured photon (but not the whole
$\rho_{AB}$) unchanged.

As seen in Fig.\ \ref{f1}, the new probabilities $p'_{\pm}$ cover all possible values from $0$ to $1$
as the measurement angle $\phi$ is varied. For $p>q$, $p'_+$ ($p'_-)$ stays above (below) the initial
value $p$ for $\phi\geq 0$, and the opposite behavior takes place for $\phi\leq 0$.
 The purities of the  states (\ref{rhoac}) are given by
\begin{eqnarray} P_{A/B_\pm}&=&1-2p'_{\pm}q'_{\pm}\sin^2\theta=
1-2pq{\textstyle\frac{[1\pm\cos(\theta-\phi)][1\pm\cos(\theta+\phi)]}{4r_{\pm}^2}}\sin^2\theta\,,\label{Pac}
\end{eqnarray}
and can  then be {\it larger or smaller} than the original purity (\ref{pa}), satisfying
\begin{equation}1-{\textstyle\frac{1}{2}}\sin^2\theta\leq P_{A/B_{\pm}}\leq 1\,.\label{Paci}\end{equation}
For $P_{A/B_-}$, the upper limit is  always reached if $\phi=\pm\theta$, as in
this case a $-$ result implies with certainty a pure post-measurement state
$|\mp\theta\rangle$ in $A$, i.e., $p'_-=0$ or $1$, as verified in Fig.\ \ref{f1}.  Such result has probability
$r_{-}=q\sin^2\theta$ ($p\sin^2\theta$) if $\phi=\theta$ ($-\theta$). The same
occurs for $P_{A/B_+}$ if $\phi=\mp(\pi-\theta)$. On the other
hand, the lower limit in (\ref{Paci}) corresponds to $p'_\pm=1/2$ and can be  reached
for angles $\phi$ satisfying
\begin{equation}\tan\phi=\frac{(p-q)\sin\theta}{\cos\theta\pm 2\sqrt{pq}}\,,\label{eq2}\end{equation}
in which case $p'_{-}=1/2$ for the roots of (\ref{eq2}) $\in[0,\pi]$ while $p'_+=1/2$ for those $\in[-\pi,0]$,  as seen in
 Fig.\ \ref{f1}. Hence by
suitable measurements and results at $B$ it is always possible to have the conditional
post-measurement state at $A$  {\it pure} or also {\it ``equilibrated''}, i.e., with equal
weights of both states of the mixture.

\subsection{Average conditional purity \label{IIC}}
The {\it average} conditional purity of $A$ after the previous measurement at $B$ is given by
\begin{eqnarray}P_{A/B_\phi}&=&r_+P_{A/B_+}+r_-P_{A/B_-}\label{PAdB}\nonumber\\
&=&1-2pq\gamma\sin^2\theta\label{kk}\,,
\end{eqnarray}
where
\begin{equation}\gamma={\frac{r_+p'_+q'_++r_-p'_-q'_-}{pq}}=
{\frac{p\sin^2(\theta-\phi)+q\sin^2(\theta+\phi)}{1-[p\cos(\theta-\phi)+q\cos(\theta+\phi)]^2}}\leq 1\,.
\end{equation}
 Hence,  in contrast with $P_{A/B_{\pm}}$, $P_{A/B_\phi}$ is never lower than the original purity:
\begin{equation}P_{A/B_\phi}\geq P_A\,,\label{des}\end{equation}
in agreement with the general results of \cite{GR.14}.   Eq.\ (\ref{PAdB}) is in
fact linearly related to the measurement dependent $S_2$ conditional entropy
\cite{GR.14}, which becomes here
\begin{eqnarray} S_2(A/B_\phi)&=&r_+S_2(\rho_{A/B_+})+r_-S_2(\rho_{A/B_{-}})\nonumber\\
&=&2(1-P_{A/B_\phi})=4pq\gamma\sin^2\theta\,,\label{S2c}\end{eqnarray}
and is never greater than the original local entropy $S_2(\rho_A)=4pq\sin^2\theta$.

The difference $P_{A/B_\phi}-P_A=2pq(1-\gamma)\sin^2\theta$ is the average conditional
purity gain at $A$ due to the local measurement at $B$,  and depends on the
measurement angle $\phi$. While it always vanishes for $\phi=0$, where
$\gamma=1$, it is otherwise positive.  Its {\it maximum} is attained for
\begin{equation}\tan\phi=\frac{\tan\theta}{p-q}\,,\label{phi}\end{equation}
in which case $\gamma=\cos^2\theta$, leading to
\begin{equation} P_{A/B}\equiv\mathop{\rm Max}_{\phi}P_{A/B_\phi}
=1-2pq\sin^2\theta\cos^2\theta\,.\label{Pmax}\end{equation}
Moreover, at this point $p'_+=q'_-$ and hence,
\begin{equation}P_{A/B_+}=P_{A/B_-}=P_{A/B}\,,\end{equation}
so that the maximum average gain is attained at an angle  where the
post-measurement local purity (but not the local state) is {\it independent} of
the result of the measurement (see also Fig.\ \ref{f4} in sec.\ \ref{IIIB}).
The maximum average purity gain is thus $2pq\sin^4\theta$.

\begin{figure}[ht]
\centerline{\hspace*{-0.2cm}\scalebox{.8}{\includegraphics{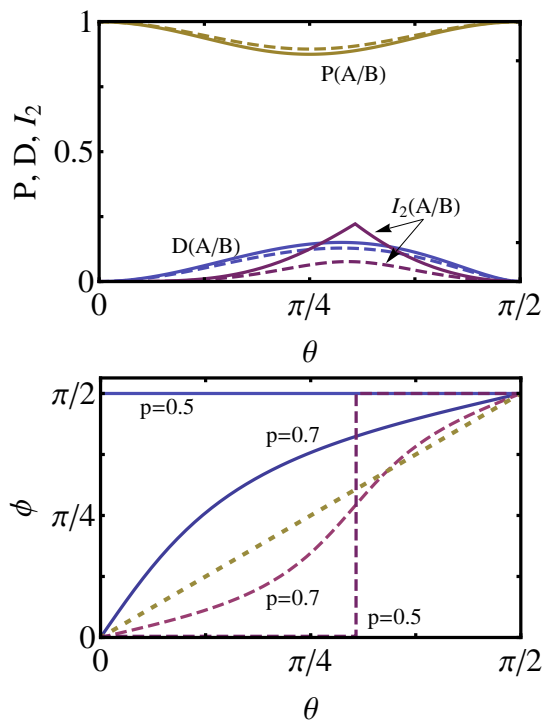}}}
\vspace{0cm} \caption{(Color online) Top: The maximum average conditional purity
$P(A/B)\equiv P_{A/B}$, Eq.\ (\ref{Pmax}), together with the quantum Discord, Eq.\ (\ref{Dmin2}) and the
minimum global purity difference, Eq.\ (\ref{I2m}), as a function of the aperture angle
$\theta$ of the state (\ref{state}) for $p=0.5$ (solid lines) and $p=0.7$ (dashed lines). Bottom: The
measurement angles which maximize the average  conditional purity  (\ref{PAdB}) (solid
lines, Eq.\ (\ref{phi}) and the global post-measurement purity (\ref{PG})
(dashed lines, Eq.\ (\ref{phi2})), as a function of the aperture angle $\theta$
for $p=0.5$ and $p=0.7$. The dotted line depicts $\theta$ for reference. For
$p=0.5$, the angle maximizing (\ref{PG}) undergoes a sharp  $0\rightarrow\pi/2$
transition at $\theta=\arccos 1/\sqrt{3}$, which originates the sharp peak in
$I_2$ seen in the top panel and which becomes smoothed out for $p>1/2$. The angle minimizing the
quantum discord (\ref{DABp}) coincides here exactly with that maximizing the conditional purity
(\ref{PAdB}) (see text).}
\label{f2}
\end{figure}

The maximum average conditional purity (\ref{Pmax}) has a deep significance.
The associated minimum $S_2$ conditional entropy
\begin{equation}S_2(A/B)=\mathop{\rm Min}_{\phi} S_2(A/B_\phi)=2(1-P_{A/B})=pq\sin^2 2\theta\,,\label{S2m}
\end{equation}
represents the {\it squared concurrence} \cite{Wo.98} between $A$ and a third system $C$ purifying
the whole system $ABC$ \cite{GR.14,KW.04,Ca.11}. $C$ can be here chosen as a
single qubit due to the  rank $2$ of $\rho_{AB}$. As a consequence (see next subsection),
the maximum average conditional purity (\ref{Pmax}) will also determine the {\it quantum discord}
of the state (\ref{state}), enabling a simple analytical expression for the latter.
The behavior of $P_{A/B}$ as a function of the ``aperture'' angle $\theta$ of the state (\ref{state})
is depicted on Fig.\ \ref{f2}. where it is seen that it reaches its maximum $1$ just for $\theta=0$
or $\pi/2$, i.e., when the state (\ref{state}) is either a product state
($\theta=0$) or a classically correlated
state ($\theta=\pi/2$), i.e., a state of zero discord in both cases.

The maximizing $\phi$ determined by (\ref{phi}) differs from $\pm
\theta$ if $\theta>0$, $pq>0$, as seen in the bottom panel of Fig.\ \ref{f2}. It satisfies $\phi\geq \theta$
if $p\geq q$, with $\phi\approx\theta+(1-p)\sin2\theta$ for $p$ close to $1$
and
$\phi\approx\frac{\pi}{2}-2(p-\frac{1}{2})/\tan\theta$ for $p$ above and close to $1/2$.
 Hence, for $p=q$ it becomes {\it independent} of $\theta$, preferring always a measurement
along the $x$ axis in the Bloch sphere (i.e., projecting onto linearly
polarized states at angles $\pm\pi/4$), as seen in Fig.\ \ref{f2}.

The previous feature is in  agreement with the
general considerations of \cite{GR.14}. The  measurement direction $\bm{k}$
in the Bloch sphere of $A$ maximizing the conditional purity of $B$ is essentially that of maximum correlation and satisfies
the generalized eigenvalue equation \cite{GR.14}
$C^TC\bm{k}=\lambda N_B\bm{k}$,
with $\lambda$ the largest eigenvalue, where $C_{\mu\nu}=\langle \sigma_{\mu}^A\otimes \sigma_\nu^B\rangle-\langle\sigma_\mu^A\rangle
 \langle\sigma^B_\nu\rangle$ is the correlation tensor of the system
and $N_B=I_3-\bm{r}_B\bm{r}_B^T$, with $\bm{r}_B=\langle \bm{\sigma}^B\rangle$
the original Bloch vector of qubit $B$. Here $\bm{r}_A=\bm{r}_B=((p-q)\sin\theta,0,\cos\theta))$ and
$C_{\mu\nu}=\delta_{\mu\nu}\delta_{\mu x}4pq\sin^2\theta\,,$
so that correlations arise just along the $x$ direction. The previous eigenvalues equation then leads to a maximizing
$\bm{k}$ in the $xz$ plane, i.e., $\bm{k}=(\sin\phi,0,\cos\phi)$, with $\phi$ satisfying Eq.\ (\ref{phi}).
And  for $p=q$,  $\bm{k}$ becomes parallel to the $x$ axis as $N_B$ becomes diagonal.

\subsection{Quantum Discord and its analytical evaluation \label{IID}}
As previously mentioned, the state (\ref{state}) has a finite quantum discord
for $\theta\in(0,\pi)$ and $pq\neq 0$. As a function of the measurement angle $\phi$,
this quantity \cite{OZ.01,HV.01}  can
be evaluated as the minimum of  the difference
\begin{equation}D(A/B_\phi)=S(A/B_\phi)-S(A/B)\,,\label{DABp}\end{equation}
where $S(A/B)=S(\rho_{AB})-S(\rho_B)=
-\sum_{\nu=\pm}(\lambda^\nu_{AB}\log_2\lambda^\nu_{AB}-\lambda^\nu_B\log_2\lambda^\nu_B)$ is
the standard von Neumann conditional entropy \cite{Wh.78} and
\begin{equation}S(A/B_\phi)=r_+S(\rho_{A/B_+})+r_-S(\rho_{A/B_-})\,,\end{equation}
the measurement dependent von Neumann conditional entropy \cite{OZ.01},
determined by the  conditional states (\ref{rhoac}) (analogous to Eq.\ (\ref{S2c})).
Hence, all quantities can  be evaluated in terms of the purities $P_{AB}$, $P_A$ and $P_{A/B_{\pm}}$
(Eqs.\ (\ref{PAB}), (\ref{pa}) and (\ref{Pac}) respectively).

The actual quantum discord $D(A/B)$ is the minimum over $\phi$ of Eq.\ (\ref{DABp}) \cite{OZ.01,Ca.08}.
The minimization should in principle be extended to general POVM
measurements, but in the present  case of a rank $2$ state,  it is
sufficient to consider just projective measurements \cite{KM.12,Ga.11}, which can here be
reduced to a measurement in the $xz$ plane.
Furthermore, the minimum of (\ref{DABp}) can be here evaluated {\it
analytically}: The minimizing measurement angle $\phi$ is {\it
exactly that which maximizes the average conditional purity},
determined by  Eq.\ (\ref{phi}), and the ensuing minimum
 {\it is a decreasing function of the maximum average
conditional purity} $P_{A/B}$, Eq.\ (\ref{Pmax}) (even though for general $\phi$,
(\ref{DABp}) is not a direct function of (\ref{PAdB})). We obtain
\begin{equation}D(A/B)=\mathop{\rm Min}_{\phi}D(A/B_{\phi})=-f_+\log_2f_+-f_-\log_2f_--S(A/B)\,,
\label{Dmin}\end{equation}
where
\begin{equation} f_\pm=\frac{1\pm\sqrt{2P_{A/B}-1}}{2}\,.\label{Dmin2}\end{equation}
Proof: According to the result of \cite{KW.04}, the minimum of $S(A/B_\phi)$ is the entanglement of
formation $E(A,C)$ between $A$ and a closing third system $C$ purifying the whole system, which
can be chosen here as a single qubit. In such a case, $E(A,C)$ is determined by the concurrence \cite{Wo.98}
$C_{AC}$ between $A$ and $C$ through  $E(A,C)=-\sum_{\nu=\pm}f_\nu\log f_\nu$,
with $f_{\pm}=(1\pm\sqrt{1-C^2_{AC}})/2$.
But $C^2_{AC}$ is just the minimum $S_2$ conditional entropy (\ref{S2m}) of $A$ given a
measurement at $B$ \cite{GR.14}, i.e. $C^2_{AB}=S_2(A/B)=2(1-P_{A/B})$, which leads to Eqs.\ (\ref{Dmin})--(\ref{Dmin2}).
We have also verified this result numerically.

For $\theta=0$ ($\rho_{AB}$ product state) or
$\theta=\pi/2$ ($\rho_{AB}$ classically correlated),  $P_{A/B}=1$ and hence $D(A/B)=0$.
Otherwise $P_{A/B}<1$ and $D(A/B)>0$, as appreciated in Fig.\ \ref{f2}. Note, however, that as a function of
$\theta$, $P_{A/B}$ is minimum at $\pi/4$ (Eq.\ (\ref{Pmax})), while $D(A/B)$ is maximum
at a slightly higher aperture  angle $\theta\approx 0.29\pi$, due to the $\theta$-dependence of the term $S(A/B)$.

\subsection{Global post-measurement purity \label{IIE}}
The average state of the whole system after the previous local measurement at $B$ is
\begin{eqnarray}\rho'_{AB}&=&r_+\,\rho_{A/B_+}\otimes\Pi_++r_-\,\rho_{A/B_-}\otimes \Pi_-\,.\label{rhop}
\end{eqnarray}
Its purity $P'_{AB}={\rm Tr}\,(\rho'_{AB})^2$ is then given by
\begin{eqnarray}P'_{AB}&=&r_+^2 P_{A/B_+}+r_-^2 P_{A/B_-}\label{PG}\\&=&
{\textstyle\frac{1}{2}}\{p\cos(\theta-\phi)+q\cos(\theta+\phi)]^2-2pq\sin^2\theta[1+\cos(\theta+\phi)
\cos(\theta-\phi)]\}\,.\nonumber
\end{eqnarray}
In contrast with Eq.\ (\ref{des}), this global post-measurement purity cannot
be greater than the original global purity (\ref{PAB}), in agreement with the general considerations of \cite{RCC.10}:
\begin{equation} P'_{AB} \leq P_{AB}\,,\end{equation}
with  $P'_{AB}<P_{AB}$ for $\theta\in(0,\pi/2)$ and $pq>0$. The
difference $P_{AB}-P'_{AB}$ is just proportional to the $S_2$ information deficit
\cite{RCC.10},
\begin{eqnarray}I_2(A,B_\phi)=S_2(\rho'_{AB})-S_2(\rho_{AB})=2(P_{AB}-P'_{AB})\,,\label{I2}\end{eqnarray}
which is always non-negative. Eq.\ (\ref{PG}) shows that it can also be evaluated just with the conditional
purities $P_{A/B_\pm}$, the initial purity $P_{AB}$ and the probabilities $r_{\pm}$.

Its minimum
\begin{equation}I_2(A,B)=\mathop{\rm Min}_{\phi}I_2(A,B_\phi)\label{I2m}\end{equation}
  which corresponds to
{maximum} global post-measurement purity $P'_{AB}$,  is proportional to the
{geometric discord} \cite{DVB.10,RCC.10,QZZ.15}.
It will then be non-zero for $\theta\in(0,\pi/2)$ and $pq\neq 0$, being maximum, like the quantum Discord,
at an angle $\theta$ above $\pi/4$, as appreciated in Fig.\ \ref{f2}.
A Renyi entropy based information deficit $I_2^R(A,B_\phi)=-\log P'_{AB}/P_{AB}$ can also be directly obtained from $I_2$ \cite{CCR.15}.

The measurement angle $\phi$ maximizing
$P'_{AB}$ (and minimizing $I_2(A,B_\phi)$)  satisfies
\begin{equation} \tan 2\phi=\frac{(p-q)\sin 2\theta}{pq+(1-pq)\cos 2\theta }\,.\label{phi2}
\end{equation}
It is not greater than that maximizing $P_{A/B_\phi}$ (Eq.\ (\ref{phi})) and can
be larger or smaller than $\theta$, with $\phi\approx \theta
-(1-p)\cos^2\theta\sin 2\theta$ for $p\rightarrow 1$. On the other hand, for
$p\rightarrow 1/2$, $\phi\rightarrow\pi/2$ just for
 $\cos\theta\leq 1/\sqrt{3}$, i.e. $\theta >\theta_c \gtrsim 0.309 \pi$, with $\phi\rightarrow 0$ for
$\cos\theta>1/\sqrt{3}$. Hence, for $p=1/2$ a {\it sharp transition} from $0$
to $\pi/2$ in the maximizing measurement angle of $P'_{AB}$, occurs at $\theta=\theta_c$ \cite{RCC.10}.
Such  sharp transition becomes smoothed out for $p>q$, as seen in Fig.\ \ref{f2}.

Since at fixed $\theta$, the minimizing  angle $\phi$ of (\ref{I2}) can differ from that
minimizing the quantum discord,
the behavior  of (\ref{I2}) as a function of the measurement angle $\phi$ may become out of phase
with that of the quantum discord,
 as will be appreciated in Fig.\ \ref{f5}. In particular, for $p=q$ and $\theta<\theta_c$, (\ref{I2})
is minimum at $\phi=0$, where $P_{A/B_\phi}$ is minimum and hence $D(A/B_\phi)$ is maximum
(as a function of $\phi$). This difference is reflecting the distinct meaning of the optimizing angles
of $P_{A/B_{\phi}}$ and $D(A/B_\phi)$ on one side, and $I_2(A,B_\phi)$ on the other side.
While the former chooses essentially the local direction associated with maximum correlation, the latter
represents the direction of a least disturbing local measurement \cite{RCC.10}, which produces the smallest
global purity decrease. Accordingly, differences can be significant for small aperture angles $\theta$, where the
latter will be closer to the $z$ axis, but will decrease as $\theta$ increases,
as seen in the bottom panel of Fig.\ \ref{f2}.
For $p=q$ they vanish in fact for $\theta>\theta_c$.

Let us finally  remark that the direction $\bm{k}$ in the Bloch sphere of $B$ of the measurement
minimizing (\ref{I2}) satisfies the standard eigenvalue equation \cite{DVB.10,RCC.10}
$(J^TJ+\bm{r}_B\bm{r}_B^T)\bm{k}=\lambda \bm{k}\label{JT}$,
with $\lambda$ the maximum eigenvalue, where $J_{\mu\nu}=\langle
\sigma_{\mu}^A\otimes\sigma_\nu^B\rangle=C_{\mu\nu}+\langle
\sigma^A_{\mu}\rangle\langle\sigma^B_{\nu}\rangle$. In the present situation
 $\bm{k}$ will lie in the $xz$ plane, i.e.,
$\bm{k}=(\sin\phi,0,\cos\phi)$, with $\phi$ satisfying Eq.\ (\ref{phi2}).
Actually, the maximizing $\phi$ is the smallest positive root of (\ref{phi2}),
the other root corresponding to the angle minimizing $P'_{AB}$ (lowest eigenvalue).

\section{Experimental verification}
\subsection{Experimental setup\label{IIIA}}

\begin{figure}[ht]
\centerline{\hspace*{0.2cm}\scalebox{.8}{\includegraphics{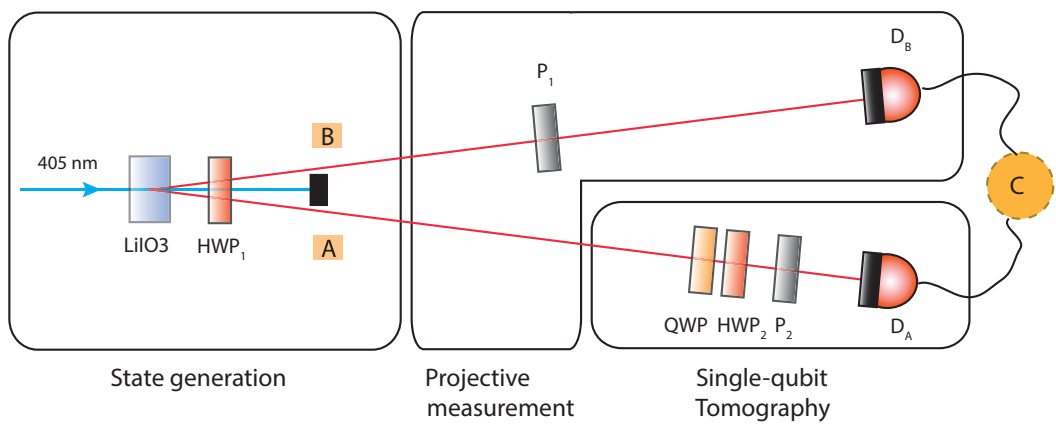}}}
\vspace{0 cm} \caption{Experimental setup used to the preparation
of a mixed polarization two-qubits state, and characterization of
the single-qubit state in $A$ conditional to a projective
measurement at $B$. QWP: quarter-wave plate; HWP: half-wave plate;
P: linear polarizer; $\mathrm{D}$: single photon detector.}
\label{fig:setup}
\end{figure}

The experimental setup is depicted in Fig.~\ref{fig:setup}. It can
be divided in three stages. In the first part, used for state
preparation, a LiIO3 nonlinear crystal cut for type-I
phase-matching, is pumped by an horizontally polarized 405nm laser
diode, that by means of spontaneous parametric down conversion
(SPDC) produces pairs of twin photons with wavelength $\lambda=810$
nm in the two-qubit polarization state $|VV\rangle$. A half-wave
plate ($\mathrm{HWP}_1$) is used to rotate the polarization of each
photon to an arbitrary direction $\theta_L$ in the laboratory
reference system, where $\theta_L=0$ corresponds to vertically
polarized photons. This angle defines the state $|\theta\rangle$ in
the Bloch sphere through the relation $\theta=2~\theta_L$. In order
to generate the mixed two-qubit states described by Eq.\
(\ref{state}), we followed the idea presented in \cite{Amselem09}
switching $\mathrm{HWP}_1$ between the two angles $\theta_L/2$ and
$-\theta_L/2$ to obtain a mixture of the two desired polarizations
with probabilities $p$ and $q$.

In the second part of the setup, a local projective polarization
measurement is done in one of the subsystems. To this end, a linear
polarizer ($\mathrm{P}_1$) in the path $B$, set at an angle $\phi_L$
or $\phi_L~+~\pi/2$ in the laboratory reference system, implements
the action of the projector $\Pi_+$ or $\Pi_-$ of Eq.\ (\ref{proj})
on the single qubit state. After that, the light is collected by the
detector $\mathrm{D}_B$. The detector is conformed by an iris that
acts as spatial filter, and an interference filter centered at 810nm
(10 nm bandwidth), followed by a lens that collects the light and
focuses it in a multi mode optical fiber coupled to a photon
counting module PerkinElmer SPCM-AQRH-13-FC. During the total time
$T$ that $\mathrm{P}_1$ is set at the angle $\phi_L$, $\mathrm{D}_B$
measures the number of single counts $n_+$, and the same is done
when $\mathrm{P}_1$ is set at the complementary angle
$\phi_L~+~\pi/2$ to register $n_-$. Then, the probability of
measuring $+$ or $-$ along the direction $\phi$ ($r_{\pm}$) given in
Eq.\ (\ref{r}) is obtained as $n_{\pm}/(n_++n_-)$.

Finally, the third part of the set up is used to perform a complete
single-qubit tomography on the subsystem $A$. An array of a
quarter-wave plate ($\mathrm{QWP}$), a half-wave plate
($\mathrm{HWP}_2$), and a linear polarizer ($\mathrm{P}_2$) in the
path of the subsystem is used to project the polarization state onto
the informational complete set of mutually unbiased basis
\cite{Ivanovic81,Wootters89}, before being detected by
$\mathrm{D}_A$. This detector consists of the same components than
$\mathrm{D}_B$, which was described above. The measurements are
performed in coincidence, using the subsystem $B$ as a trigger. In
this way it is possible to reconstruct the conditional density
matrix $\rho_{A/B=\pm}$ of the reduced state of $A$ after obtaining
the outcome $\pm$ of the projective measurement along direction
$\phi$ at $B$. Although these observables could be obtained by
performing a conditional purity measurement, the used set up
aditionally allows to verify that the post-measurement local
conditional state was of the form (\ref{rhoac}).

\subsection{Results\label{IIIB}}

\begin{figure}[ht]
\centerline{\hspace*{0.5cm}\scalebox{.8}{\includegraphics{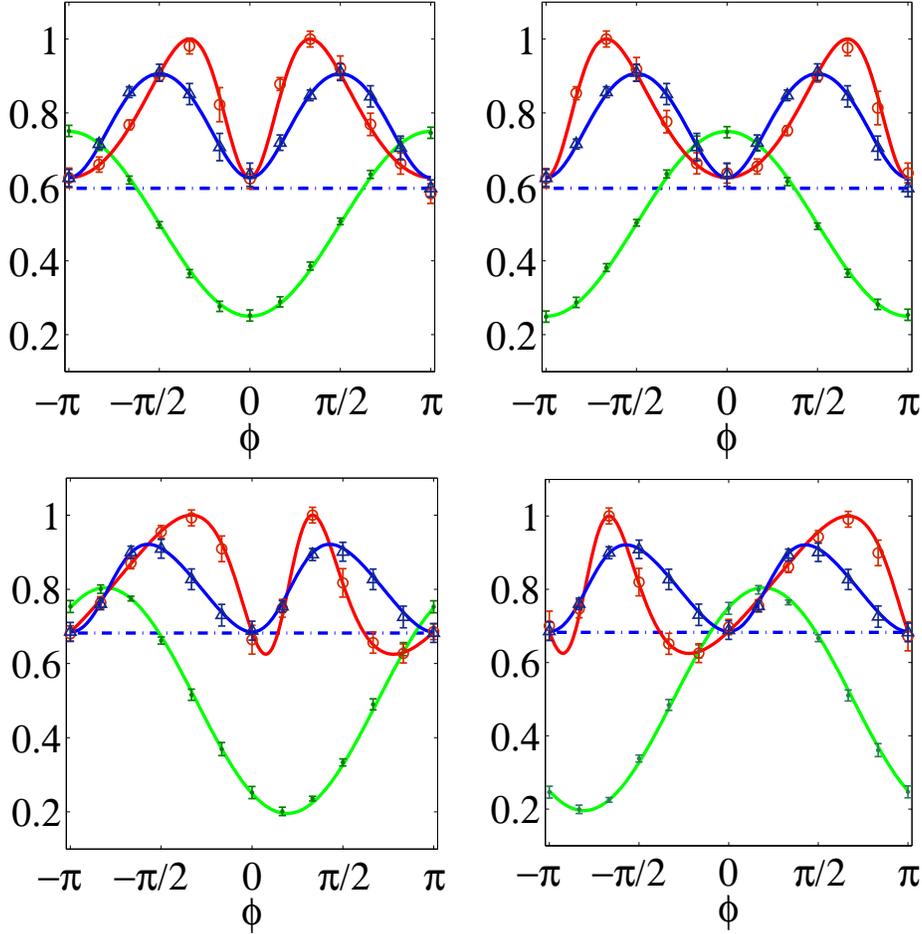}}}
\vspace{0cm}\caption{(Color online) Experimental results obtained from each
projective measurement described by $\Pi_\pm$. It shows the purity
of the state $A$ after obtaining the result $-$ (left panels) or $+$
(right panels) at $B$, $P_{A/B_{\pm}}$ (Eq.\ (\ref{Pac})) (red
circles), and the probability to obtain this result, $r_{\pm}$
(Eq.\ (\ref{r})) (green points). Additionally, each graphic shows the average
conditional purity $P_{A/B_\phi}$, Eq.\  (\ref{PAdB}) (blue
triangles). In all cases the initial state is a two-qubit state with
$\theta=\pi/3$. For the top panels $p$=0.5 and for the bottom panels
$p$=0.7. Solid lines correspond to the theoretical values for these
measures as a function of $\phi$. The dashed line indicates the
measured value of the local purity $P_{A}$ in the initial state
before conditional measurement. }\label{f4}
\end{figure}

\begin{figure}[ht]
\centerline{\hspace*{0.5cm}\scalebox{0.7}{\includegraphics{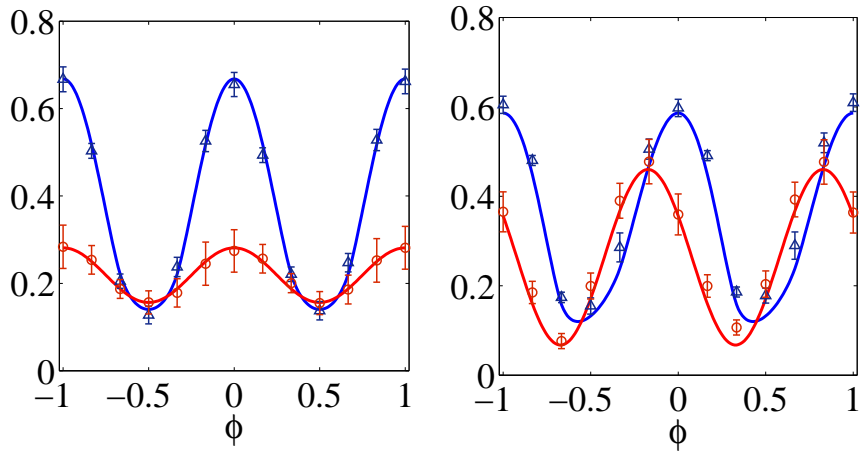}}}
\vspace{-.25 cm}
\caption{(Color online) Experimental results obtained from each
projective measurement described by $\Pi_\pm$. The minimum of the
blue line (triangles) is the value of the quantum discord while the
minimum of the red line (circles) is the value of geometric
discord (minimum global purity difference). The initial state is a
two qubit state corresponding to $\theta=\pi/3$ with $p$=0.5 (left
panel) and $p$=0.7 (right panel). Solid lines correspond to the
theoretical values for these measures as a function of the
measurement angle $\phi$.}\label{f5}
\end{figure}

As mentioned above, the first section of the experimental setup
generates a two-photon field in the polarization state given by
Eq.~(\ref{state}). In order to validate the generation process, a
previous complete tomography \cite{Kwiat01} for different mixed
two-qubit states was done. Afterwards, maximum likelihood technique
(ML) was applied to obtain the best state estimation consistent with
the requirements of a physical state \cite{Fiurasek01}. We
quantified the quality of the preparation process by means of the
fidelity $F\equiv
\mathrm{Tr}\left(\sqrt{\sqrt{\sigma}\rho\sqrt{\sigma}}\right)$
between the density matrix of the state intended to be prepared,
$\sigma$, and the density matrix of the state actually prepared and
reconstructed by tomography, $\rho$. In all cases, fidelities $F >
0.98$ for the initial state $\rho_{AB}$ were obtained. After this
previous characterization we performed the projective measurements
and the conditional one-qubit tomography implemented in the second
and third parts of the setup (Fig. \ref{fig:setup}). The post
measurement state $\rho_{A/B_{\pm}}$ in Eq. (\ref{rhoac}) was
obtained after applying the ML technique to the experimental
results. The plots in Fig. \ref{f4} show the probabilities for the
projective measurements $r_{\pm}$ and the conditional purities
$\mathrm{Tr} \rho^2_{A/B_{\pm}}$ for two different initial states
together with the theoretical predictions given by Eqs. (\ref{r})
and (\ref{Pac}). Additionally, we plot the average conditional
purity $P_{A/B_\phi}$ which was obtained from the experimental
results as $r_+P_{A/B_+}+r_-P_{A/B_-}$ and theoretically as
$1-2pq\gamma\sin^2\theta$ (Eq. (\ref{kk})). In Fig. \ref{f5} we plot
the experimental and theoretical results for $D(A/B)$ and
$I_2(A,B_{\phi})$, whose minimum values as a function of the
projective measurement $\Pi_\phi$ correspond to the quantum discord
and the geometric discord, respectively. As follows from Sec.
\ref{IID} and Sec. \ref{IIE} both quantities can be evaluated from
the purities $P_{AB}$, $P_{A}$, and $P_{A/B_\pm}$. For this purpose,
the density matrix of the reduced state $\rho_A$ was obtained in a
similar way to $\rho_{A/B_{\pm}}$ but considering the single counts
in $D_A$ without taking in account the results in $D_B$.

\section{Conclusions\label{IV}}
We have examined in full detail the quantum correlation properties
of the two-parameter mixed states (\ref{state}), characterized by an
aperture angle $\theta$ and a probability or weight $p$. Such states arise
naturally as reduced two-spin states of spin 1/2 chains, either at
the mean field approximation level or exactly in the vicinity of the
factorizing field, and can be easily generated by photonic qubit
states in a linear-optics architecture. We have derived simple exact
analytical expressions for the conditional purity of one of the
qubits after a local measurement on the other qubit, including its
maximum average. These quantities allow one to also determine quantum
correlation measures.

In particular, we have derived a simple exact analytical
expression for the quantum discord of the state, which unveils its
direct connection with the previous maximum average
conditional purity, valid for the present states. 
This result enables a straightforward experimental evaluation of the quantum discord
of the state through a conditional single photon purity measurement. 
Such determination can assess, for instance, its potential for 
quantum protocols such as quantum state merging. The global
post-measurement purity and the associated information deficit $I_2$ were as well analytically
evaluated and compared with the previous measures. The experimental
results showed a remarkable agreement with the theoretical
predictions. The analysis indicates that the form of the reduced state
of the unmeasured photon $A$ remains unchanged after a remote
measurement on  photon $B$, although selection of the measurement
angle allows full control of the probabilities characterizing this
reduced state, including pure and maximally mixed limits. At the
same time, due to the non-orthogonality of the states involved these
probabilities do affect the eigenstates of the conditional reduced
states, and any local measurement does affect the average
post-measurement global state. Possible applications to cryptography
and metrology are currently under investigation.

\ack{The authors acknowledge support from  CIC (RR) and
CONICET(LR,JJMV,NG,NC,CI,SL) of Argentina. L.R. thanks the Institut
f\"ur Gravitationsphysik and Leibniz Universit\"at Hannover for kind
hospitality. This work was supported by UBACyT 20020130100727BA,
CONICET PIP 112201101-00902, CONICET PIP 112200801-03047, and ANPCYT
PICT 201002179 (Argentina).}

\end{document}